\documentclass{article}\usepackage[utf8]{inputenc}
\usepackage{float}
\usepackage{amsmath}
\usepackage{amssymb}
\usepackage{esvect}
\usepackage[letterpaper, margin=0.7 in]{geometry}
\usepackage{authblk}
  \usepackage{amsmath}
    \usepackage{amssymb}
    \usepackage{amsthm}
    \usepackage{amsfonts}
    \usepackage{braket}
\usepackage{amsmath}
\usepackage{amsfonts}
\usepackage{amssymb}
\usepackage{hyperref}
\usepackage{graphicx}
\usepackage{caption}
\usepackage{subcaption}
\usepackage{float}

\title{Noncommutative  scalar fields in compact spaces:
quantisation and implications
}
\author{Mir Mehedi Faruk$^{1,2}$\thanks{Corresponding author: muturza3.1416@gmail.com, mir.faruk15@imperial.ac.uk, mir.mehedi.faruk@cern.ch, mir.faruk@mail.mcgill.ca}}
\author{Mishkat Al Alvi$^{3,4}$}
\author{Wasif Ahmed$^{5,6}$}
\author{Md Muktadir Rahman$^{3,7}$}
\author{Arup Barua Apu$^{8}$}

\affil{
Bose Centre for Advanced Study and Research in Natural Science, University of Dhaka, Bangladesh. $^1 $\\
Physics Department, McGill University
Montreal, QC H3A 2T8, Canada$^2$\\
Department of Theoretical Physics, University of Dhaka, Bangladesh$^3$\\
Department of Physics and Astronomy, University of Minnesota,
Duluth, Minnesota 55812. USA$^4$.\\
Department of Physics, University of Dhaka, Bangladesh$^5$\\
International Centre for Theoretical Physics, Strada Costiera 11, Trieste 34151 Italy$^6$\\
Department of Physics, University of South Dakota, 414 E. Clark St., Vermillion, SD 57069. USA$^7$.\\
Department of Physics, Western Illinois University, 1 University circle,  Macomb, IL, 61455, USA$^8$\\
}

\begin{document}
\maketitle
\begin{abstract} 
 In this paper we consider a two component scalar field theory, with noncommutativity in its conjugate momentum space.   
We quantize such a theory in a compact space with the help of dressing transformations and 
we reveal
a significant effect of introducing such noncommutativity
as the splitting of the energy levels of each
individual mode that constitutes the whole system.
We further compute the thermal partition function exactly with predicted deformed dispersion relations  from noncommutative theories and compare the results with usual results.
It is found that
 thermodynamic quantities in noncommutative  models, irrespective of whether the model is more deformed in infrared/UV region, show deviation from standard results in high temperature region.

\end{abstract}
\section{Introduction}
It is  
a common concept that the usual picture of spacetime as a
smooth pseudo-Riemannian manifold would break down
due to  quantum gravity effects
at very short distances of the order of the Planck length.
The deviation from the flat-space
concept at  the order of the Planck
length is actually motivated from
 new concepts such as quantum groups\cite{book1},
quantum loop gravity\cite{loop}, deformation theories\cite{mag1}, noncommutative
geometry\cite{ncgeo}, string theory\cite{string1} etc.
Besides, the
idea of noncommutative spacetime was also discovered in
string theory and in the matrix model of M theory, where
in the  certain limit due to
the presence of a background field B,
noncommutative gauge theory appears\cite{string1}. 
Recently an increasing interest
towards noncommutative theories has been triggered by the results in
string theory\cite{string2}. A vast number
of papers dealing with the problem of formulating noncommutative  field theory \cite{1,2,3,4,5,6,7,8,9,bal,khelili,ir} has appeared in literature. 
Some implications of such noncommutativity in field theory including
 connections between Lorentz  invariance violation    and  noncommutativity of fields\cite{8}, deformed energy eigenvalue of the  Hamiltonian\cite{8,bal}, replusive Casimir force\cite{khelili},
 deformed Kac-Moody Algebra\cite{bal}, Bose Einstein condensation\cite{bec}, 
 noncommutative field gas driven inflation\cite{nccosmo},
 UV/IR mixing\cite{sp1}, GZK cutoff\cite{bal1}, path integral \cite{sp2}, matter-antimatter asymmetry\cite{8} etc have appeared in literature.
 Due to the huge potential of noncommutative field theories to produce interesting results, such theories 
need  extensive attention.
 \\ \\
Several groups have reported that quantum
gravity relics could be seen from the Lorentz violating dispersion relations\cite{cam}.
Lorentz invariance
is then considered as a good low-energy symmetry which
may be violated at very high energies.
As our
low-energy theories are quantum field theories (QFT),
it is interesting to explore  possible generalizations
of the QFT framework which could produce
departures from exact Lorentz symmetry.
The
assumption of noncommutativity in the field
space of a QFT produces Lorentz-violating dispersion
relations in both compact\cite{bal} and non-compact space\cite{khelili}.
We expect that signatures of noncommutativity will appear in future 
experiments involving 
ultrahigh-energy cosmic rays,
the cosmic microwave background
(CMB),  or  neutrino experiments\cite{9}.
As the blackbody spectrum
coincides with 
CMB radiation  with
great accuracy, at least for low to medium frequencies,
the deformed blackbody radiation (with deformed dispersion relation) needs 
to be well studied.
Blackbody radiation with deformed dispersion coming from other theories such as doubly special relativity\cite{mir1}, generalized uncertainty principle\cite{gup}, phenomenological quantum gravity theories\cite{mir2,cam}
etc. are  already well studied. But the blackbody radiation with deformed dispersion relation
due to noncommutativity in field theory is yet to be studied in detail.
Although Balachandran et. al.\cite{bal} has visited some of the properties of deformed backbody radiation, they did not solve the partition function which we set to do in this paper.
\\\\
Noncommutativity in QFT  has been studied heavily by several groups
as
any deviation from the
usual free massless boson theory may have some influence
on the modeling of  experimental observables. In usual two component  scalar field theory
the field $\phi^i$ $(i=1,$ $2)$ 
and the canonical conjugate momentum $\pi^i$ are assumed to be operators
satisfying the canonical commutation relations,
\begin{subequations}
\begin{align}
 [{\phi}^i(\vv{x},t),{\phi}^j(\vv{y},t)] &= 0, \\
    [{\pi}_i(\vv{x},t), {\pi}_j(\vv{y},t)]     &= 0 ,\\
    [{\phi}^i(\vv{x},t),{\pi}_j(\vv{y},t)]  &= i\delta^i_j\delta(\vv{x}-\vv{y}).
\end{align}
\end{subequations}
Here, $(\vv{x},t)$ are elements of base space.
In their work, Balachandran et. al.\cite{bal}
and Khelili\cite{khelili} considered  noncommutative massless  scalar fields
with
commutative base space and noncommutative target space.
Therefore, the above commutation relations 
take the form\footnote{We have used hat in field space noncommutativity and tilde for momentum space noncommutativity.},
\begin{subequations}\label{eq:12}
\begin{align}
 [\hat{\phi}^i(\vv{x},t),\hat{\phi}^j(\vv{y},t)] &= i
  \epsilon^{ij} \theta\delta(
    \vv{x}- \vv{y}), \\
    [\hat{\pi}_i(\vv{x},t), \hat{\pi}_j(\vv{y},t)]     &= 0, \\
    [\hat{\phi}^i(\vv{x},t),\hat{\pi}_j(\vv{y},t)]  &= i\delta^i_j\delta(\vv{x}-\vv{y}),
\end{align}
\end{subequations}
where  $\epsilon^{ij}$ is
an antisymmetric constant matrix and
$\theta$ is a parameter with the dimension of length.
After constructing the Hamiltonian formulation
of this theory and quantizing it in a compact space,
Balachandran et. al. have obtained a splitting
 of the energy levels of each
individual mode that constitutes the whole system. 
The resemblance of this effect to the well known
Zeeman effect in a quantum system in the presence
of a magnetic field is  noticed\cite{bal}.
Balachandran et. al. have considered a $S^1\times S^1 \times S^1 \times \mathbb{R}$ type geometry, where 
all compactified spatial
coordinates, i.e., $S^1$, are of same radius.
In this manuscript we  investigate a different type of noncommutativity in  scalars. 
Here, we explore the case where the fields are commutative but the conjugate momentum space is noncommutative.
Therefore the commutation relations are of the form,
\begin{subequations}\label{eq:12}
\begin{align}
 [\tilde{\phi}^i(\vv{x},t),\tilde{\phi}^j(\vv{y},t)] &= 0, \\
    [\tilde{\pi}_i(\vv{x},t), \tilde{\pi}_j(\vv{y},t)]     &= i
  \epsilon^{ij} \theta\delta(
    \vv{x}- \vv{y}), \\
    [\tilde{\phi}^i(\vv{x},t),\tilde{\pi}_j(\vv{y},t)]  &= i\delta^i_j\delta(\vv{x}-\vv{y}).
\end{align}
\end{subequations}
At first, we canonically quantize 
a free massless boson theory with the commutation relation 
in Eq. 3 in $(1+1)$ dimension following the regularization procedure shown in the seminal paper of Balachandran et. al.\cite{bal}. 
We then  construct a Fock space, since the Hamiltonian
 can be diagonalized using the Schwinger representation
of SU(2).
Afterwards, we generalize the results in arbitrary dimensions.
Finally, we  compute the thermal partition function for the deformed Hamiltonian due to noncommutativity in Eq. (2) and (3) and compare them. 
\section{Review on commutative  scalar field theory in compact space }
Let us consider a theory in a ($d+1$)-dimensional base space 
and the target space is set in a commutative plane $\mathbb{R}^2$. 
Here, we present a free massless boson theory with
commutative base space and noncommutative target space.
The target space is the space where the field take its
values. The spatial part
of the base space is a $d$ dimensional torus.
Now if the field components are denoted by $\phi^i$ where $i = 1,2$ then we can write,
\begin{subequations}\label{eq:1}
\begin{align}
    &&\phi : S_1^1 \times S_2^1 \times ... \times S_d^1 \times \mathbb{R} &\longrightarrow \mathbb{R}^2, \\
  &&  (\vv{x},t) &\longmapsto \phi (\vv{x},t).
\end{align}
\end{subequations}
We invoke that each spatial direction is compactified 
in $S^1_j$ with radius $R_j$, which causes the field components to be periodic in the spatial coordinates,
\begin{align}
\phi(\vv{x}+\vv{R},t)=\phi(\vv{x},t)
\end{align}
Let us write the components of the field $\phi^i (\vv{x},t)$ as a Fourier series,
\begin{equation}\label{eq:2}
    \phi^j (\vv{x},t) = \sum\limits_{\vv{n},j} e^{-\frac{2 \pi i}{R_j} \vv{n}.\vv{x}} \varphi^j_{\vv{n}}(t),
\end{equation}
where, $\vv{n} = (n_1, n_2, ... n_d)$ and thus the Fourier components of the field are,
\begin{equation}\label{eq:3}
   \varphi^j_{\vv{n}}(t) = \frac{1}{M} \sum\limits_{j} \int d^dx\ e^{\frac{2 \pi i}{R_j} \vv{n}.\vv{x}} \phi^i (\vv{x},t).
\end{equation}
Here, \begin{equation}
    V=\prod_{j=1} ^{d} R_j,
\end{equation}
One can notice a real condition $\phi _n ^*(t)=\phi_{-n}(t)$
from eq. (7). 
Now the Lagrangian is,
\begin{equation}\label{eq:4}
    L = \frac{1}{2} \sum\limits_j \int d^dx \big[ (\partial_t \phi^j)^2 - (\nabla \phi^j)^2  \big].
\end{equation}
The above Lagrangian in terms of Fourier mode for $d$ dimensional target space is
\begin{equation}\label{eq:5}
    L = \frac{V}{2} \sum \limits_{i,\vv{n}} \Bigg\{ \dot{ \varphi^i}_{\vv{n}} \dot{ \varphi^i}_{-\vv{n}} - \omega_{\vv{n}}^2  \varphi^i_{\vv{n}} \varphi^i_{-\vv{n}} \Bigg\}.
\end{equation}
Here, $\omega_{\vv{n}}$ is defined as,
\begin{equation}
\omega_{\vv{n}}^2=(2\pi)^2\sum_{j=1}^d (\frac{n_j^2}{R_j^2}).
\end{equation}
From the Lagrangian, we can now evaluate the expression for the momentum,
\begin{subequations}\label{eq:6}
\begin{align}
    \pi^i_{\vv{n}} &= \frac{\partial L}{\partial \dot{\varphi}^i_{\vv{n}}} \\              
                   &= V\dot{\varphi}^i_{-\vv{n}}.
\end{align}
\end{subequations}
Armed with the Lagrangian and the momentum, we can finally write the Hamiltonian of the system,
\begin{subequations}\label{eq:7}
\begin{align}
    H &= \sum \limits_{i,\vv{n}} \pi^i_{\vv{n}} \dot{\varphi}^i_{\vv{n}} - L \\
      &= \sum\limits_{i,\vv{n}} \Big[ \frac{1}{2V} \pi_{\vv{n}}^i \pi_{-\vv{n}}^i + \frac{V}{2}\omega_{\vv{n}}^2 \varphi_{\vv{n}}^i \varphi_{-\vv{n}}^i \Big].
\end{align}
\end{subequations}
So,  $\omega_{\vv{n}}$ corresponds to the frequencies of the set of harmonic oscillators describing the system as defined in eq (11).
\section{Non-commutative field theory}
It is well known that the 
phase space of a single particle in $\mathbb{R}^2$ has a
natural group structure
which is the semidirect product of $\mathbb{R}^2$
with $\mathbb{R}^2$. The generators of its Lie algebra can be taken to
be coordinates $x^a$, with $a = 1, 2$, and momenta being $p_a$.
\begin{subequations}\label{eq:8}
\begin{align}
[x^a, x^b]=0,\\
[x^a, p_b]=i\delta^a_b,\\
[p_a,p_b]=0.
\end{align}
\end{subequations}
One can now twist/deform the generators of the above
algebra into $x^a$, $p^ b$ and thus obtain  new algebras\cite{bal}.
\subsection{Model 1: Noncommutativity in momentum space, Quantisation in (1+1) Dimension}
In this model, the
algebra of derivatives is deformed, but the function
algebra is not. Here we twist (or deform) the generators of the above
algebra into $\tilde{x^a}, \tilde{p^ b}$ and thus obtain a new algebra. 
\begin{subequations}\label{eq:9}
\begin{align}
[\tilde{x}^{a},\tilde{x}^{b}] &= 0 \\
    [\tilde{p}^{a},\tilde{p}^{b}] &= i\epsilon^{ab}{\theta} = i\tilde{\theta}^{ab} \\    
    [\tilde{x}^{a},\tilde{p}^{b}] &= i\delta^{a}_{b}
\end{align}
\end{subequations}
Here,  $\theta$ is a parameter and
$\epsilon^{ab}$ is
an antisymmetric constant matrix.
We can relate the non-commutative coordinates with their commutative counterparts in terms of the deformation parameter with the help of dressing transformation\cite{bal,dressing1,dressing2,dressing3},
\begin{subequations}\label{eq:11}
\begin{align}
    \tilde{p}^{a} &= p^{b} + \frac{1}{2} \tilde{\theta}^{ab}  x_{b}, \\
    \tilde{x}^{b} &= x^{b}.
\end{align}
\end{subequations}
The above dressing transformation map (16) can be easily generalized to
 scalar field theory. 
We start with a free real massless bosonic  field. Its base
space is a cylinder with circumference $R$ and its target
space is $\mathbb{R}^2$.
\begin{eqnarray}
    &&\phi : S^1 \times \mathbb{R} \longrightarrow \mathbb{R}^2\\
    &&(x,t) \longrightarrow \phi(x,t)
\end{eqnarray}
The compactification of the space coordinate makes each
field component periodic, i.e.
\begin{eqnarray}
\phi^i(x+R,t)=\phi^i(x,t).
\end{eqnarray}
As a result  $\phi^i(x,t)$ can be written as  Fourier expansion as in Eq. (6), where the Fourier components can be rewritten as eq (7). Now following the spirit of Balachandran et. al.\cite{bal}, we rewrite the eq. (3),
\begin{subequations}\label{eq:12}
\begin{align}
 [\tilde{\phi}^i(\vv{x},t),\tilde{\phi}^j(\vv{y},t)] &= 0, \\
    [\tilde{\pi}_i(\vv{x},t), \tilde{\pi}_j(\vv{y},t)]     &= i
  \epsilon^{ij} \theta(\sigma;
    \vv{x}- \vv{y}), \\
    [\tilde{\phi}^i(\vv{x},t),\tilde{\pi}_j(\vv{y},t)]  &= i\delta^i_j\delta(\vv{x}-\vv{y}),
\end{align}
\end{subequations}
where,
\begin{eqnarray}
\theta(\sigma;
    \vv{x}- \vv{y})=\frac{\theta}{\sigma\sqrt{2\pi}}e^{\frac{-(x-y)^2}{2\sigma^2}}.    
\end{eqnarray}
The redefinition of the term $\theta \delta(x-y)$ is to be seen as a
regularization procedure (see reference \cite{bal}).
It should be noted that these new commutation
relations reduce to those in eq.  (3) in the limit  $\sigma\rightarrow 0$.
And both in the limit
$\theta \rightarrow 0$ we obtain the usual commutation relation.
The  parameter $\sigma$ indicates a new
distance scale in the equal time commutation relations
for the fields \cite{bal}. Here,
$\sigma$ has dimension of length, and $\theta$ has dimension\footnote{in $d+1$ dimension $\theta$ has dimension of (length)$^{2d-1}$, but $\sigma$ always have dimension of length.} of
(length)$^{1}$.
The novelty of the  Balachandran et al's work was to regularise the delta function in the commutation
relation, which prevented the energy density to diverge with respect to frequency.
We will use their method in our work as well.
Following eq (16), the dressing transformation for field theory in this model is,
\begin{eqnarray}
    &&\tilde{\pi}^i (\vv{x},t) =\pi^i (\vv{x},t)+\frac{1}{2}\epsilon^{ij}\int dy \theta (\sigma;
    \vv{x}- \vv{y}) 
    \phi_j(\vv{y},t),\\
  &&  \tilde\phi^i(\vv{y},t)= \phi^i(\vv{y},t).
\end{eqnarray}
Here, $i=1,2$
and $\tilde{\pi}(\vv{x},t) $ is the  
canonical conjugate momentum of the field $\tilde{\phi}(\vv{x},t)$. 
The map defined above reads in Fourier modes,
\begin{eqnarray}
    \tilde{\pi}^i_n=\pi^i_n+\frac{1}{2R}\epsilon^{ij}\varphi^j_{-n}\theta(n),  \\
        \tilde{\varphi}^i_n=\varphi^i_n. 
\end{eqnarray}
The original commutation relations are,
\begin{subequations}\label{eq:12}
\begin{align}
     [\varphi^i_m, \varphi^j_n]=[\pi^i_m,\phi^j_n]=0, \\
        [\varphi^i_m, \pi^j_n]=i\delta_{mn}\delta^{ij},
\end{align}
\end{subequations}
and the modified commutation relations  in Fourier space are,
\begin{subequations}\label{eq:12}
\begin{align}
[\tilde{\varphi^i_m}, \tilde{\varphi^j_n}]=0, \\
    [\tilde{\pi}^i_n,\tilde{\pi}^j_m]=\frac{i}{R}\epsilon^{ij}\theta(n)\delta_{n+m,0}, \\
[\tilde{\varphi}^i_n, \tilde{\pi}^j_m]=i\delta_{nm}\delta^{ij}.
\end{align}
\end{subequations}
Considering free massless noncommutative  
scalar fields, the Lagrangian can be written as,
\begin{equation}
    L= \frac{1}{2}\sum_i\int dx[ (\partial_t \tilde\phi^i)^2 - (\partial_x \tilde\phi^i)^2].
\end{equation}
Now, the Hamiltonian in Fourier space,
\begin{align}
\tilde{H} &=\sum_{i}\frac{(\tilde{\pi}^i_0)^2}{2R}+\frac{1}{2R}\sum_{i, n\neq0}  \nonumber 
\left \{\tilde{\pi}^i_n\tilde{\pi}^i_{-n}+(2\pi|n|g)^2\tilde{\varphi}^i_n \tilde{\varphi}^i_{-n} \right \} \\ \nonumber
&=\frac{1}{2R}\sum_i ( \pi^i_0\pi^i_0+\frac{1}{R}\theta(0)\epsilon^{ij}\varphi^j_0\pi^i_0+\frac{1}{4R^2}\theta^2(0)\varphi^i_0\varphi^i_0) + \\ & \frac{1}{2R}\sum_{i, n\neq 0}\big \{ \pi^i_{n}\pi^i_{-n}+[(2\pi|n|)^2+\frac{\theta^2(n)}{4R^2}]\varphi^i_{n}\varphi^i_{-n}+\frac{1}{R}\theta(n)\epsilon^{ij}\varphi^j_n\pi^i_n \big \} 
\end{align}
Now, the standard harmonic oscillator Hamiltonian can be written as,
\begin{equation}
H_n=\sum_i\big (\frac{1}{2M}\pi^i_n\pi^i_{-n}+\frac{1}{2}M\bar{\omega}_n^2\varphi^i_n\varphi^i_{-n} \big )
\end{equation}
Comparing with the above equation we can write,
\begin{equation}
\bar{\omega}_n=\frac{1}{M}\sqrt{(2\pi|n|)^2+\frac{\theta^2(n)}{4R^2}}.
\end{equation}
Now, we can define the creation and annihilation operators as,
\begin{equation}
a^i_n=\sqrt{\frac{\Delta_n}{2}}(\varphi^i_n+i\frac{\pi^i_{-n}}{\Delta_n}),
\end{equation}
\begin{equation}
a^{i\dagger}_n=\sqrt{\frac{\Delta_n}{2}}(\varphi^i_{-n}-i\frac{\pi^i_n}{\Delta_n}).
\end{equation}
Here,
\begin{equation}
\Delta_n=R\bar{\omega}_n=\sqrt{(2\pi|n|g)^2+\frac{\theta^2(n)}{4R^2}}.
\end{equation}
Now,
\begin{equation}
[a^i_m, a^j_n]=[a^{i\dagger}_m, a^{j\dagger}_n]=0,
\end{equation}
\begin{equation}
[a^i_m, a^{j\dagger}_n]=\delta_{mn}\delta^{ij}.
\end{equation}
The original Hamiltonian can be written in terms of the creation and annihilation operators defined above,
\begin{align}
H_n &=\sum_i\big (\frac{1}{2R}\pi^i_n\pi^i_{-n}+\frac{R}{2}\bar{\omega}^2_n\varphi^i_n\varphi^i_{-n}\big ) \\ 
&=\sum_i\frac{1}{2}\bar{\omega}_n\big (a^i_n a^{i\dagger}_n+ a^{i\dagger}_{-n}a^i_{-n}\big ) 
\end{align}
Now, after normal ordering the Hamiltonian looks like
\begin{equation}
H_n=\sum_i\bar{\omega}_n a^{i\dagger}_n a^i_n.
\end{equation}
Now, the $\varphi^j_n\pi^i_n$ term of the Hamiltonian can be written as
\begin{align*}
\epsilon^{ij}\varphi^j_n\pi^i_n &= \frac{-i}{2}\epsilon^{ij}[a^{j\dagger}_{-n}a^i_{-n}-a^j_na^{i\dagger}_n] \\
&= i\epsilon^{ij}a^{i\dagger}_na^j_n. \\
\end{align*}
Normal ordering has been used to derive this expression. The complete Hamiltonian can be written as,
\begin{equation}
\tilde{H}=H_0+\sum_{i,n\neq 0}\bar{\omega}_na^{i\dagger}_na^i_n+\frac{i}{2}\frac{\theta(n)}{M^2}\sum_{i,j; n\neq0}\epsilon^{ij}a^{i\dagger}_na^j_n.
\end{equation}
Now lets define new creation and annihilation operators,
\begin{equation}
A^1_n=\frac{1}{\sqrt{2}}(a^1_n-ia^2_n),
\end{equation}
\begin{equation}
A^2_n=\frac{1}{\sqrt{2}}(a^1_n+ia^2_n).
\end{equation}\\
So, if we write the Hamiltonian in terms of these new creation and annihilation operators, it looks like\footnote{following ref. \cite{bal} we ignore the zero mode. It is not relevant, since it is
associated with the overall translation of the system.},
\begin{align}
\tilde{H} &= H_0+\sum_n\bar{\omega}_n[A^{1\dagger}_nA^1_n+A^{2\dagger}_nA^2_n]-\frac{g}{2M^2}\sum_n\theta(n)[A^{1\dagger}_nA^1_n-A^{2\dagger}_nA^2_n] \\
&= H_0+\sum_n(\bar{\omega}_n-\frac{1}{2R^2}\theta(n))A^{1\dagger}_{n}A^1_n+\sum_n(\bar{\omega}_n+\frac{1}{2R^2}\theta(n))A^{2\dagger}_{n}A^2_n. 
\end{align}
This is how energy splitting occurs due to noncommutativity in momentum space.
A  consequence of such noncommutativity in momentum space is the appearance
of a term proportional to a component of angular
momentum in the Hamiltonian of the theory. It affects
the splitting of the energy levels. Splitting is also noticed if noncommutativity is introduced in field space \cite{bal}. But the functional form of the two types of splitting are quite different.
\subsubsection{ The deformed conformal generators}
Now, we will have a look at the deformed conformal generators. The deformed Hamiltonian 
written with hatted operators is,
\begin{equation}
\tilde{H}=\sum_i\frac{(\tilde{\pi}^i_0)^2}{2R}+\frac{1}{2R}\sum_{i, n\neq0}\big \{\tilde{\pi}^i_n\tilde{\pi}^i_{-n}+(2\pi|n|)^2\tilde{\varphi}^i_n\tilde{\varphi}^i_{-n} \big \}.
\end{equation}
The deformed creation and annihilation operators can be written as,
\begin{equation}
\tilde{a}^i_n=\frac{1}{\sqrt{4\pi|n|}}(2\pi |n|\tilde{\varphi}^i_n+i\tilde{\pi}^i_{-n}),
\end{equation}
\begin{equation}
\tilde{a}^{i\dagger}_{n}=\frac{1}{\sqrt{4\pi|n|}}(2\pi |n|\tilde{\varphi}^i_{-n}-i\tilde{\pi}^i_{n}).
\end{equation}
So, 
\begin{equation}
[\tilde{a}^i_m,\tilde{a}^j_n]=\frac{1}{4\pi|n|}\frac{-i}{R}\epsilon^{ij}\theta(n)\delta_{n+m,0},
\end{equation}
\begin{equation}
[\tilde{a}^i_m,\tilde{a}^{j\dagger}_n]=\delta^{ij}\delta_{mn}+\frac{1}{4\pi|n|}\frac{i}{R}\epsilon^{ij}\theta(n)\delta_{mn},
\end{equation}
\begin{equation}
[\tilde{a}^{i\dagger}_m,\tilde{a}^{j\dagger}_n]=\frac{1}{4\pi|n|}\frac{-i}{R}\epsilon^{ij}\theta(n)\delta_{n+m,0}.
\end{equation}
It should be noted that, if we make $\theta\rightarrow0$, the creation-annihilation operators of noncommutative theories coincide with the usual theory.
The generators of the modified $U(1)$ Kac-Moody algebra would be, \\\\
For $n>0$,
\begin{equation}
J^i_n=-i\sqrt{n}\tilde{a}^i_n
\end{equation}
\begin{equation}
\bar{J}^i_n=-i\sqrt{n}\tilde{a}^i_{-n}
\end{equation}
For $n<0$,
\begin{equation}
J^i_n=i\sqrt{-n}\tilde{a}^{i\dagger}_{-n}
\end{equation}
\begin{equation}
\bar{J}^i_n=i\sqrt{-n}\tilde{a}^{i\dagger}_n
\end{equation}
The commutators between the generators can be written as,
\begin{equation}
[J^i_m, J^j_n]=m\delta^{ij}\delta_{n+m,0}+\frac{i}{4\pi R}\epsilon^{ij}\theta(n)\delta_{n+m,0}
\end{equation}

\begin{equation}
[\bar{J}^i_n,\bar{J}^j_m]=m\delta^{ij}\delta_{n+m,0}+\frac{1}{4\pi }\frac{i}{R}\epsilon^{ij}\theta(n)\delta_{n+m,0}
\end{equation}

\begin{equation}
[J^i_m,\bar{J}^j_n]=\frac{1}{4\pi }\frac{i}{R}\epsilon^{ij}\theta(n)\delta_{n,m}
\end{equation}
It can be easily observed that
a term dependent upon noncommutative parameter $\theta$
has appeared 
 in
the commutation relations of the $U(1)$ Kac-Moody algebra. This deformed $U(1)$
Kac-Moody due to momentum space noncommutativity is quite different compared to field space noncommutativity deformed  $U(1)$
Kac-Moody\cite{bal}.
Now, the non-zero mode terms of the Hamiltonian can be written as,
\begin{equation}
\frac{2\pi}{R}\sum_{i,n>0}(J^i_{-n}J^i_n+\bar{J}^i_{-n}\bar{J}^i_n)
\end{equation}

So,
\begin{equation}
[\tilde{H}, J^k_{-m}]=\frac{2\pi}{R}\sum_{i,m>0}\big \{2mJ^k_{-m}+\frac{1}{2\pi g}\frac{i}{R}\theta(n)\epsilon^{ik} (J^i_{-m}+\bar{J}^i_m) \big \}
\end{equation}

Now we can write the conformal generators,
\begin{equation}
\hat{L}_0=\frac{1}{2}\sum_i {J^i_0}^2+\sum_{i, n>0}J^i_{-n}J^i_n
\end{equation}

\begin{equation}
\hat{L}_n=\frac{1}{2}\sum_{i,m, n\neq0}J^i_{n-m}J^i_m
\end{equation}

\begin{equation}
\hat{\bar{L}}_0=\frac{1}{2}\sum_i{\hat{\bar{J}}^i_0}^2+\sum_{i,n>0}\bar{J}^i_{-n}\bar{J}^i_n
\end{equation}

\begin{equation}
\hat{\bar{L}}_n=\frac{1}{2}\sum_{i,m}\bar{J}^i_{n-m}\bar{J}^i_m
\end{equation}

Here,
\begin{equation}
J^i_0=\bar{J}^i_0=\frac{1}{\sqrt{4\pi }}\sqrt{\pi^i_0\pi^i_0+\frac{1}{R}\theta(0)\epsilon^{ij}\varphi^j_0\pi^i_0+\frac{1}{4R^2}\theta^2(0)\varphi^i_0\varphi^i_0}
\end{equation}

So, the Hamiltonian can be written as,
\begin{equation}
\tilde{H}=\frac{2\pi}{R}(\hat{L}_0+\hat{\bar{L}}_0)
\end{equation}

\subsection{Momentum noncommutativity in (d+1) Dimension}
We will now generalize the results of the noncommutativity in (d+1) dimension.
The spatial part
of the base space is a $d$ dimensional torus, just like eq. 4 but with non commutative  scalar field. 
Therefore, the field can be written as,

\begin{align}
\tilde{\varphi}^i(\vec{x},t) &= \sum_{\vec{n}}exp[2\pi i(\frac{n_1x_1}{R_1}+\frac{n_2x_2}{R_2}+\cdots+\frac{n_dx_d}{R_d})]\tilde{\varphi}^i_{\vec{n}}(t) \\
&= \sum_{\vec{n}} exp[2\pi i(\sum_j \frac{n_jx_j}{R_j})]\tilde{\varphi}^i_{\vec{n}}(t)
\end{align}
The Fourier components can be written as,
\begin{equation}
\tilde{\varphi}^i_{\vec{n}}(t)=\frac{1}{R_1R_2\cdots R_d}\int d^dx \; exp[-2\pi i(\sum_j\frac{n_jx_j}{R_j})]\tilde{\varphi}^i(\vec{x},t) 
\end{equation}
So, the Lagrangian looks like
\begin{equation}
L=\frac{1}{2}\sum_i\int d^dx \; [(\partial_t\tilde{\varphi}^i)^2-(\mathbf{\nabla}\tilde{\varphi}^i)^2] \nonumber
\end{equation}
If we write Lagrangian in terms of Fourier modes, then
\begin{equation}
L=\frac{1}{2}R_1R_2\cdots R_d\sum_{i,\vec{n}}\big \{ \dot{\tilde{\varphi}}^i_{\vec{n}}\dot{\tilde{\varphi}}^i_{\vec{-n}} - 4\pi^2[\sum_j\frac{n_j^2}{R_j^2}]\tilde{\varphi}^i_{\vec{n}}\tilde{\varphi}^i_{\vec{-n}} \big \}
\end{equation}
The canonical momentum is defined by:
\begin{equation}
\tilde{\pi}^i_{\vec{n}}=\frac{\partial L}{\partial \dot{\tilde{\varphi}}^i_{\vec{n}}}=R_1R_2\cdots R_d  {\dot{\tilde{\varphi}}}^i_{\vec{-n}}
\end{equation}
Now, The deformation map can be written as:
\begin{equation}
\tilde{\varphi}^i_n=\varphi^i_n
\end{equation}
\begin{equation}
\tilde{\pi}^i_n=\pi^i_n+\frac{1}{2R_1R_2\cdots R_d}\epsilon^{ab}\varphi^j_{-n}\theta(\vec{n})
\end{equation}
The commutation relationships between the deformed field modes are:
\begin{equation}
[\tilde{\varphi}^i_m,\tilde{\varphi}^j_n]=0
\end{equation}
\begin{equation}
[\tilde{\pi}^i_m,\tilde{\pi}^j_n]=\frac{i}{R_1R_2\cdots R_d}\epsilon^{ij}\theta(\vec{n})\delta_{n+m,0}
\end{equation}
\begin{equation}
[\tilde{\varphi}^a_n, \tilde{\pi}^b_m]=i\delta_{mn}\delta^{ab}
\end{equation}
Here the term $\theta(\vec{n})$ is defined as,
\begin{equation}
\theta(\vec{n})=\theta exp[-2\pi^2\sigma^2\sum_j\frac{n_j^2}{R_j^2}]
\end{equation}
If we write the Hamiltonian in terms of the hatted operators then,
\begin{equation}
H=\frac{1}{2R_1R_2\cdots R_d}\sum_i \tilde{\pi}^i_0\tilde{
}{\pi}^i_0+\frac{1}{2R_1R_2\cdots R_d}\sum_{i,\vec{n}\neq0}\big \{ \tilde{\pi}^i_n\tilde{\pi}^i_{\vec{-n}}+(2\pi R_1R_2\cdots R_d)^2[\sum_j\frac{n_j^2}{R_j^2}]\tilde{\varphi}^i_{\vec{n}}\tilde{\varphi}^i_{\vec{-n}} \big \}
\end{equation} 
So, using the dressing transformation, 
\begin{align}
H &= H_0+\frac{1}{2R_1R_2\cdots R_d}\sum_{i,\vec{n}\neq 0}\pi^i_{\vec{n}}\pi^i_{-\vec{n}}+ \frac{1}{2R_1R_2\cdots R_d}\sum_{i,\vec{n}\neq 0} \big \{ (2\pi R_1R_2\cdots R_d)^2 [\sum_j\frac{n_j^2}{R_j^2}] \nonumber \\ &+ \theta^2(\vec{n})\frac{1}{4R_1^2R_2^2\cdots R_d^2} \big \}\varphi^i_{\vec{n}}\varphi^i_{\vec{-n}}+ \frac{\theta(\vec{n})}{2R_1R_2\cdots R_d}\sum_{i,\vec{n}\neq0}\frac{1}{R_1R_2\cdots R_d}\epsilon^{ij}\pi^i_n\varphi^j_n 
\end{align}
Now, let us define,
\begin{align}
H_{\vec{n}} &= \sum_i\frac{1}{2R_1R_2\cdots R_d}\pi^i_{\vec{n}}\pi^i_{\vec{-n}}+ \sum_i\frac{1}{2R_1R_2\cdots R_d}[(2\pi R_1R_2\cdots R_d)^2 \big \{\sum_j\frac{n_j^2}{R_j^2} \big \} \nonumber\\ &+ \theta^2(\vec{n})\frac{1}{4R_1^2R_2^2\cdots R_d^2} ]\varphi^i_{\vec{n}}\varphi^i_{\vec{-n}}  \\ 
\end{align}
The standard harmonic oscillator Hamiltonian can be written as,
\begin{equation}
H_{\vec{n}}=\sum_i(\frac{1}{2V}\pi^i_{\vec{n}}\pi^i_{\vec{-n}}+\frac{V}{2}\bar{\omega}_n^2\varphi^i_{\vec{n}}\varphi^i_{\vec{-n}})
\end{equation}
Comparing these two equations we can write
\begin{equation}
V=R_1R_2\cdots R_d
\end{equation}
\begin{equation}
\bar{\omega}_n=\frac{1}{M}\sqrt{(2\pi R_1R_2\cdots R_d)^2\big \{\sum_j\frac{n_j^2}{R_j^2} \big \} + \frac{\theta^2(\vec{n})}{4R_1^2R_2^2\cdots R_d^2}}
\end{equation}
Let us now define creation and annihilation operators,
\begin{equation}
a^i_{\vec{n}}=\sqrt{\frac{\Delta_{\vec{n}}}{2}}(\varphi^i_{\vec{n}}+i\frac{\pi^i_{\vec{-n}}}{\Delta_{\vec{n}}})
\end{equation}
\begin{equation}
{a^i_{\vec{n}}}^{\dagger}=\sqrt{\frac{\Delta_{\vec{n}}}{2}}(\varphi^i_{\vec{n}}-i\frac{\pi^i_{\vec{n}}}{\Delta_{\vec{n}}})
\end{equation}
Here,
\begin{equation}
\Delta_{\vec{n}}=V\bar{\omega}_{\vec{n}}=\sqrt{(2\pi R_1R_2\cdots R_d)^2\big \{\sum_j\frac{n_j^2}{R_j^2} \big \} + \frac{\theta^2(\vec{n})}{4R_1^2R_2^2\cdots R_d^2}}
\end{equation}
Using these operators, the Hamiltonian $H_{\vec{n}}$ can be written as:
\begin{equation}
H_{\vec{n}}=\sum_i\frac{\omega_{\vec{n}}}{2}(a^i_{\vec{n}}{a^i_{\vec{n}}}^{\dagger}+(a^i_{-\vec{n}}{a^i_{-\vec{n}}}^{\dagger})
\end{equation}
The time-ordered form of this Hamiltonian is,
\begin{equation}
\sum_{\vec{n}\neq 0}H_{\vec{n}}=\sum_{i,\vec{n}\neq 0}\bar{\omega}_{\vec{n}}({a^i_{\vec{n}}}^{\dagger}a^i_{\vec{n}})
\end{equation}
The last term of the Hamiltonian can be written as,
\begin{equation}
\frac{\theta(\vec{n})}{2R_1R_2\cdots R_d}\sum_{i,\vec{n}\neq 0}\frac{i}{R_1R_2\cdots R_d}\epsilon^{ij}{a^i_{\vec{n}}}^{\dagger}a^j_{\vec{n}}
\end{equation}
So, the total Hamiltonian can be written as,
\begin{align*}
H&=H_0+\sum_{i,\vec{n}\neq 0}\bar{\omega}_{\vec{n}}{a^i_{\vec{n}}}^{\dagger}a^i_{\vec{n}}+\frac{i}{2}\frac{\theta(\vec{n})}{V^2}\sum_{i,j,\vec{n}\neq 0}\epsilon^{ij}{a^i_{\vec{n}}}^{\dagger}a^j_{\vec{n}} \\
\end{align*}
We can repeat the Schwinger process done previously and define new creation and annihilation operators, $A^1_{\vec{n}}$ and $A^2_{\vec{n}}$. Using these operators, the Hamiltonian,
\begin{equation}
H=H_0+\sum_{\vec{n}\neq 0}\big \{ (\bar{\omega}_{\vec{n}}-\frac{1}{2V^2}\theta(\vec{n})){A^1_{\vec{n}}}^{\dagger}A^1_{\vec{n}}+ (\bar{\omega}_{\vec{n}}+\frac{1}{2V^2}\theta(\vec{n})){A^2_{\vec{n}}}^{\dagger}A^2_{\vec{n}} \big \}
\end{equation}
So, the  energy levels can be read as,
\begin{equation}
\Lambda^1_{\vv{n}}=\bar{\omega}_{\vec{n}}-\frac{1}{2V^2}\theta(\vec{n})
\end{equation}

\begin{equation}
\Lambda^2_{\vv{n}}=\bar{\omega}_{\vec{n}}+\frac{1}{2V^2}\theta(\vec{n})
\end{equation}
\subsection
{Model 2: Noncommutativity  in field space, quantisation in (d+1) Dimension}
One can also consider noncommutativity in field space instead of the momentum space. Canonical  
quantisation of such theories in compact space has already taken under consideration by Bal $et.$ $al$. The twisted algebra reads in this model, 
\begin{subequations}\label{eq:9}
\begin{align}
    [\hat{x}^{a},\hat{x}^{b}] &= i\epsilon^{ab}\bar{\theta} = i\bar{\theta}^{ab} \\    [\hat{p}^{a},\hat{p}^{b}] &= 0 \\
    [\hat{x}^{a},\hat{p}^{b}] &= i\delta^{a}_{b}
\end{align}
\end{subequations}
Also, the corresponding equal time commutation relations in field theory are,
\begin{subequations}\label{eq:12}
\begin{align}
 [\hat{\phi}^i(\vv{x},t),\hat{\phi}^j(\vv{y},t)] &= i
  \epsilon^{ij} \theta(\sigma;
    \vv{x}- \vv{y})  \\
    [\hat{\pi}_i(\vv{x},t), \hat{\pi}_j(\vv{y},t)]     &= 0 \\
    [\hat{\phi}^i(\vv{x},t),\hat{\pi}_j(\vv{y},t)]  &= i\delta^i_j\delta(\vv{x}-\vv{y})
\end{align}
\end{subequations}
To see the the canonical quantisation procedure
in compact space 
of this type of model see ref.\cite{bal}.
The spatial part of the base space is a  d 
dimensional torus. But in their paper,
they
considered compactified spatial
coordinates, i.e., $S^1$, all of them with the same radius $R$. But here we present the 
results
 invoking  that each spatial direction is compactified 
in $S^1$ with radius $R_j$.
Mimicking the calculation of Bal $et$ $al.$\cite{bal}, we find out the quantized  Hamiltonian (normal ordered),
\begin{eqnarray}
H=H_0+ \sum_{n \neq 0}\omega_{\vec{n}} \{  \Gamma^1_{\vv{n}}  {A^1_{\vv{n}}} ^{\dagger}  A^1_{\vv{n}}  +   \Gamma^2_{\vv{n}}  {A^2_{\vv{n}}}^{\dagger}  A^2_{\vv{n}} \}
\end{eqnarray}
where,
\begin{eqnarray}
 &&\Gamma^1_{\vv{n}} = \Omega_{\vv{n}} - \frac{\omega_{\vv{n}}\theta(n)}{2} \\
 &&   \Gamma^2_{\vv{n}}= \Omega_{\vv{n}} + \frac{\omega_{\vv{n}}\theta(n)}{2} \\
 &&       \theta(\vv{n}) = \theta exp[-2 \pi^2 \sigma^2 (\frac{n_1 ^2}{R_1^2}  +\frac{n_2 ^2}{R_2^2}  +....   +\frac{n_d^2}{R_d^2}  )]\\
&&     \omega_{\vv{n}}^2 =4\pi^2(\frac{n_1 ^2}{R_1^2}  +\frac{n_2 ^2}{R_2^2}  +....   +\frac{n_d^2}{R_d^2}  )\\ 
&& \Omega_{\vv{n}}^2=1+\pi^2  \theta(\vv{n})^2 (\frac{n_1 ^2}{R_1^2}  +\frac{n_2 ^2}{R_2^2}  +....   +\frac{n_d^2}{R_d^2}  )
\end{eqnarray}
In the limit, $R_1=R_2=...=R_d$ the above equations coincides with the result of Balachandran et. al.\cite{bal}. Therefore the splitted deformed dispersion relation take the form below,
\begin{eqnarray}
 \Lambda^1_{\vv{n}} = \omega_{\vv{n}}(\Omega_{\vv{n}} - \frac{\omega_{\vv{n}}\theta(n)}{2})\\
 \Lambda^2_{\vv{n}} = \omega_{\vv{n}}(\Omega_{\vv{n}} + \frac{\omega_{\vv{n}}\theta(n)}{2})
\end{eqnarray}

\section{Deformed Blackbody Radiation}
In this section, we analyze the blackbody radiation due to 
deformed energy momentum relation coming from field space (eq. 101 and 102) and momentum space  noncommutativity  (eq. 91 and 92). Although Balachandran et. al. has briefly discussed it, they did not calculate the thermodynamic quantities. However, we have numerically evaluated the thermodynamic quantities and we will present them in this section.
We start from the partition function of quantum gases in grand canonical ensemble \cite{path},
\begin{equation}
Z=Tr( e^{-\beta H}).
\end{equation}
As there is a split in energy eigenvalues due to both types of noncommutativity,
we find out from above eq. \cite{bal,path},
\begin{eqnarray}
\ln Z=- \sum_{\vv{k}\neq 0}(\ln (1 -e^{-\beta \Lambda^1_{\vv{k}}})+\ln (1 -e^{-\beta \Lambda^2 _{\vv{k}}})).
\end{eqnarray}
 Here, 
 $\Lambda^1$ and $\Lambda^2$
 refer to two distinct classes of modes due to noncommutativity conditions,
 $\vv{k}$ is the momentum vector
 and $\beta=\frac{1}{T}$.\footnote{we have chosen Boltzmann constant $k_B=1$, $c=1$}
Therefore, the internal energy,
\begin{equation}
U=-\frac{\partial}{\partial \beta}lnZ.
\end{equation}
Now, the entropy $S$ can be obtained from the partition function,
\begin{eqnarray}
 S=-\frac{\partial F}{\partial T},
\end{eqnarray}
where $F=\frac{1}{\beta} \ln Z$ is the free energy.
\begin{figure}[H]
\begin{subfigure}{0.5\textwidth}
  \centering
  \includegraphics[width=0.98\linewidth]{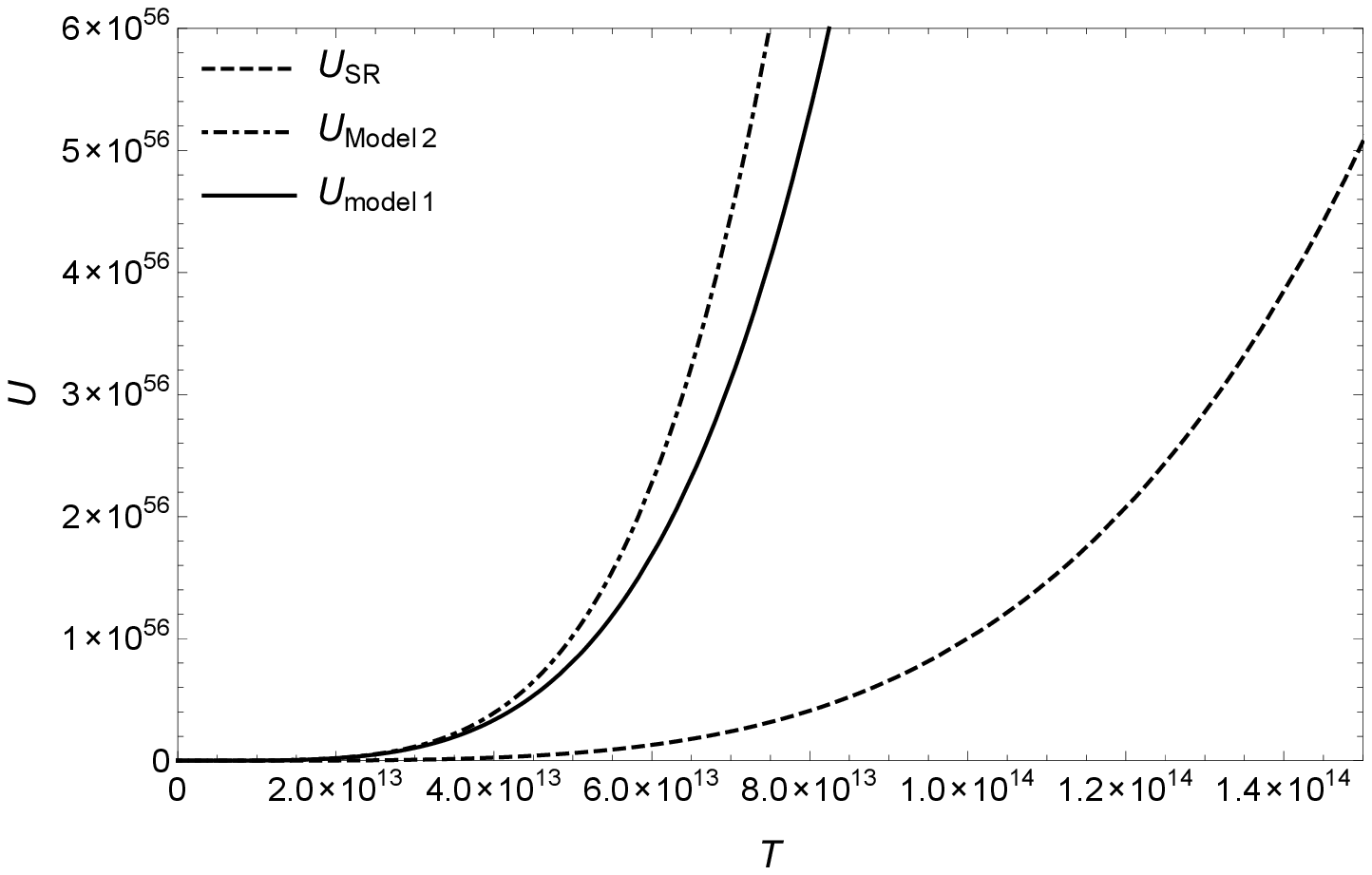}
  \caption{Internal energy $U$, versus temperature}
  \label{fig:sub1}
\end{subfigure}%
\begin{subfigure}{.5\textwidth}
  \centering
  \includegraphics[width=0.98\linewidth]{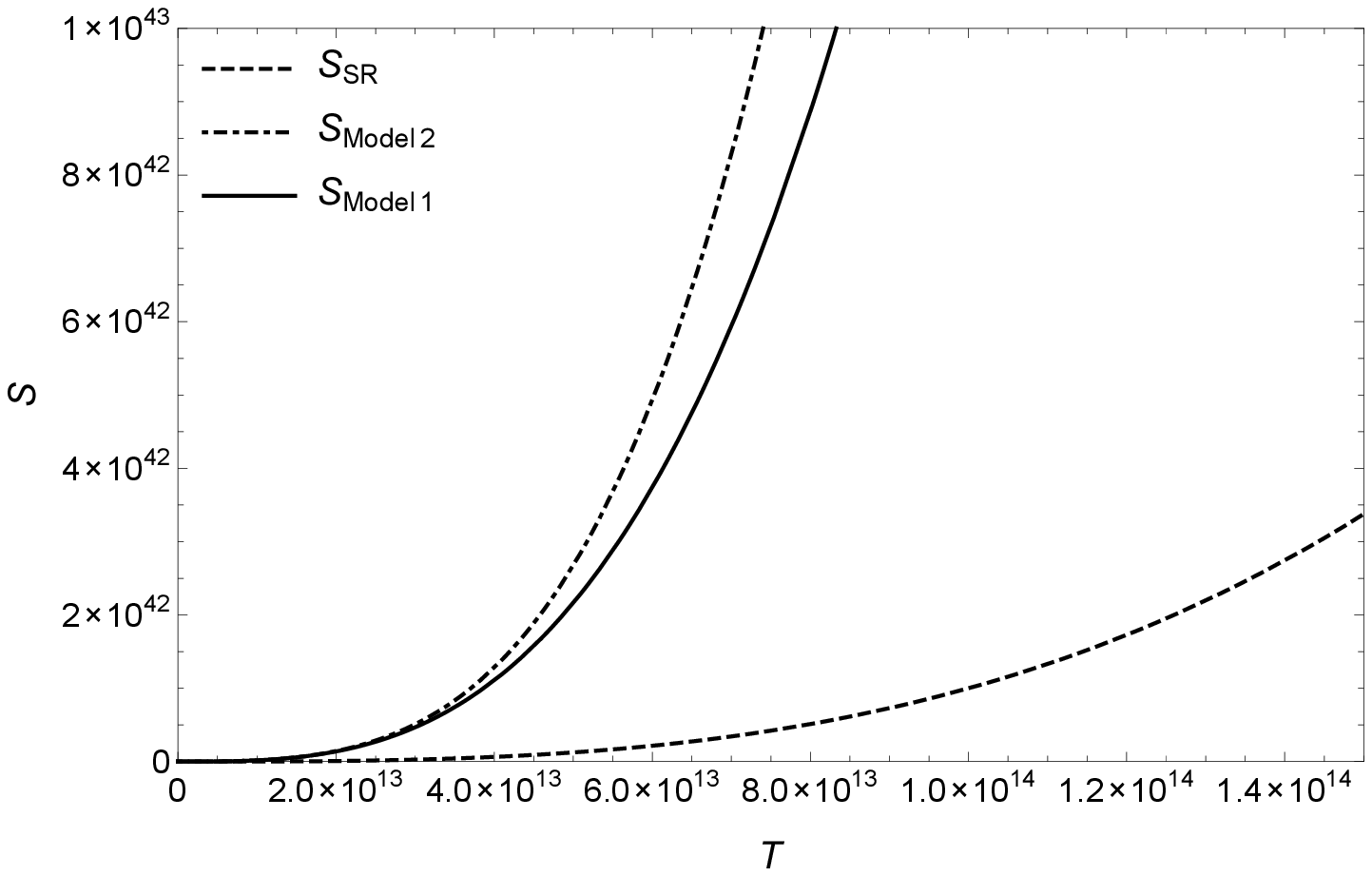}
  \caption{ Entropy $S$, versus temperature }
  \label{fig:sub2}
\end{subfigure}
\caption{
Plot of internal energy $U$ (figure a) and entropy $S$ (figure b) of blackbody radiation against temperature
$T$ for  the special relativity theory and the noncommutative  models. Following the ref.  \cite{8}, \cite{bal}, \cite{nccosmo}  we have chosen $\theta=1.7\times 10^{-14} $ and $\sigma= 10^{-15}$.}
\label{fig:test}
\end{figure}
In the thermodynamic limit we consider all $R_i \rightarrow \infty$
which allows us to convert the sum in eq. (104) to an integral.
We performed the integrals numerically using mathematica\cite{math}, choosing specific values for  $\theta$ and $\sigma$ following Balachandran et. al\cite{bal}. 
Finally, using eq (105) and (106) we evaluate the internal energy and entropy for both type of noncommutative models and usual special relativity (SR). We have compared the results in fig. 1. We have found
that   all the three models agree in the lower temperature
but the noncommutativity effects surely modify 
the Stefan-Boltzmann law ($U\propto T^4$), which is clearly visible in the high temperature regime. 
But at high temperature  a significant difference is noticed between them (see figure 1a). In the high temperature regime it is seen that at any temperature $T=T_0$ the internal energy coming from these models maintain a relation $U_{SR} (T_0)<U_{model1} (T_0)<U_{model2} (T_0) $.
This trend is also noticed for other thermodynamic quantities such as entropy, specific heat etc.
The dispersion relation predicted from both types of noncommutative filed theories  are clearly Lorentz violating.
This trend of faster rate of growth (with respect to temperature) of thermodynamic quantities at a high temperature regime 
compared to SR 
is also noticed in 
other Lorentz violating studies on the thermodynamics of blackbody radiation\cite{cam}. 
As one can see 
in the model $1$ the modifications of the dispersion relations $(91)$ and $(92)$ occur for small wave number $n$ 
 and become the usual ones in the large
$n$ limit (more deformed in infrared region)
whereas in
model $2$ the dispersion relations $(101)$ and $(102)$ have large modifications for large wave number $n$
and becomes the usual one in small $n$ limit
(more deformed in UV region). But
interestingly, thermodynamic quantities in both of the models show deviation from standard special relativity results in high temperature regime.
\\\\\
In case of noncommutative
models, due to Lorentz violating dispersion relations, number of available states grow. As a result when we do the
integration over all the modes in eq. (104) we find the modified internal energy. The Planck distribution function picks
up a smaller value in low temperature region compared to high temperature region. The noncommutative
parameter makes some modification in Planck distribution but it is not extremely drastic. As a result when we do
integration over it no such significant change is noticed in low temperature due to noncommutative parameter. Now as
the temperature rises abruptly the Planck distribution attends higher values and a small change due to noncommutative
parameter makes the change big enough that when do integration over all the modes a difference is noticed. The effect
of noncommutative parameter in internal energy is less clear in low temperature region, as in lower temperature region the
(modified) Planck
distribution picks up a very small value.
As a result thermodynamic quantities in both of these models, irrespective of whether the model is more deformed in infrared/UV region, show deviation from standard results in high temperature regime.
\section{Conclusion}
In this paper we have canonically quantized a noncommutative  scalar field theory in a compact space
with noncommutativity in momentum space following the seminal work of Balachandran et. al.\cite{bal}. 
As a result of this noncommutativity, we have noticed
the splitting of the energy levels of each
individual mode that constitutes the whole system. This type of splitting in energy eigenvalue was also noticed in noncommutative
 scalars, where noncommutativity is in field space\cite{bal}. But the functional forms of deformed dispersion relations due to two types of noncommutativity 
are quite different and as a result their prediction are also quite different in blackbody radiation.
We have paid special attention to
the special case of 1+1 dimensional theory and found out the deformed conformal generators. 
We are in a process to evaluate the deformed Virasoro algebra for noncommutative theories and find out the status of central charge in such field theories.
The  central charge is a very significant concept in conformal theories as the theories are characterized by this number.
A different central charge would imply a new interpretation of central extension. 
Such noncommutativity would be even more interesting for gauge fields as the cancellation of degrees of freedom with  Gupta-Bleuler quantization
or Faddev-Popov method by appearance of ghost fields which can lead to new physics.
The central charge of the ghost fields play a significant role in 
the critical dimension of string theory.
Furthermore, it should be noticed that we have considered  the spatial part
of the base space is a $d$ dimensional torus
where we invoked that each spatial direction is compactified 
in $S^1_j$ with radius $R_j$. The reason behind keeping the result more general is we are in a process to compute 
Casimir force for noncommutative theories in compact space. The general result  will help  us to make any particular direction, say $R_1$ to keep finite and other direction  to put in the  bulk limit.
In the future, we would also like to 
investigate
the finite temperature status of these type of theories. As a result  we notice that, thermdynamic quantities in both of these models show deviation from standard result  
\section{Acknowledgements}
One of the authors (MMF) would thanks the  Bose Centre for Advanced Study and Research in Natural Sciences, University of Dhaka, Bangladesh for financial support. The authors would also
like to thank Professor Dr. Arshad Momen for the fruitful discussions. The authors express their gratitude to Baptiste Ravina and Liana Islam for their help to present the manuscript.
We  sincerely thank the referees for their fruitful comments.  

\end{document}